*Machine learning assisted multiscale modeling of composite phase change materials for Li-ion batteries' thermal management*

Felix Kolodziejczyk*[a], Bohayra Mortazavi**[b], Timon Rabczuk[c] and Xiaoying Zhuang[b,c]

[a]Institute of Continuum Mechanics, Leibniz Universität Hannover, Appelstraße 11, 30157 Hannover, Germany.
[b]Chair of Computational Science and Simulation Technology, Institute of Photonics, Department of Mathematics and Physics, Leibniz Universität Hannover, Appelstraße 11,30157 Hannover, Germany.
[c]College of Civil Engineering, Department of Geotechnical Engineering, Tongji University, Shanghai, China.

Abstract

In this work, we develop a combined convolutional neural networks (CNNs) and finite element method (FEM) to examine the effective thermal properties of composite phase change materials (CPCMs) consisting of paraffin and copper foam. In this approach, first the CPCM microstructures are modeled using FEM and next the image dataset with corresponding thermal properties is created. The image dataset is subsequently used to train and test the CNN's performance, which is then compared with the performance of a popular network architecture for image classification tasks. The predicted thermal properties are employed to define the properties of the CPCM material of a battery pack. The heat generation and electrochemical response of a Li-ion cell during the charging/discharging is simulated by applying Newman's battery model. Thermal management is achieved by the latent heat of paraffin, with copper foam for enhancing the thermal conductivity. The multiscale model is finally developed using FEM to investigate the effectiveness of the thermal management of the battery pack. In these models the thermal properties estimated by the FEM and the CNN are employed to define the CPCM materials properties of a battery pack. Our results confirm that the model developed on the basis of a CNN can evaluate the effectiveness of the battery pack's thermal management system with an excellent accuracy in comparison with the original FEM models.

1. Introduction

Addressing the global climate change is one of the biggest and most important challenges of today's society. With around 23%, the vehicular transportation sector contributes a big part to human-made, energy-related greenhouse gas emissions[1]. Therefore, full electrification of the automotive sector is an inevitable task. Hybrid electric vehicles (HEVs) and electric vehicles (EVs) are the industry's first steps towards an electric future. Currently, rechargeable batteries play a critical role as the energy storage system of HEVs and EVs. Particularly, lithium-ion batteries (LiB) have to date remained unbeaten in terms of reaching energy efficiencies of above 90% [2]. The most challenges for an efficient electric vehicle (BEV) are nonetheless associated within the battery system. There are numerous challenges for current LiB technologies such as need for thermal management due to self-heating or hot and cold climates, the risk of thermal runaway due to overheating or traffic accidents, capacity loss due to aging and low energy densities compared to conventional fuels, the expensiveness of lithium and cobalt metals used in electrodes and other related ace moral and environmental problems.

As electrochemical performance and the safety of a battery is heavily influenced by the temperature, a thermal management system is needed to keep the temperature in the optimal range[3,4]. Heat is produced by the battery itself during cycling due to electrochemical processes, short-circuiting or physical damage [5]. High temperatures can lead to higher rates of self-discharge, loss of the capacity, aging of the battery and shortening of life, thermal runaway and even explosion of the module [2–4,6,7]. When scaling up batteries for high power applications, such as an EV, thermal management becomes among the main concerns [8]. Thermal management systems use different mediums to achieve heating or cooling, such as the ambient air, a heat transfer fluid, an insulator, a phase change material (PCM) or a combination of the previous [9]. The thermal management system can be active or passive. An active system is a vehicle-powered source of heat or cold, while a passive system relies on the ambient environment [9]. As active cooling showed results of improving battery performance by 30-40%, it introduces complexity into vehicle design and battery operation control [6,8,10]. Further, conventional passive air or active liquid cooling systems are bulky and fans and pumps used in an active design drains battery power, which reduces the effectively usable energy and therefore vehicle range [6,9].

As an efficient and low cost thermal management system, PCMs achieve cooling due to its high latent heat, which can be easily placed between the cells and acts as a heat sink. When the temperature exceeds the PCM's melting point, all further heat is used for the phase change until the melting is finished and therefore preventing the battery temperature to rise [4]. The stored latent heat is released during the solidifying process, which is an important advantage when operating EVs in cold climates [8]. Paraffin waxes are often chosen for the design of PCMs, as they are cheap, show high heat capacity with low density and have melting temperatures within a battery's operational range [8,11]. However, paraffin's poor thermal conductivity (~0.2 W/mK) limits the maximum cooling power and therefore also the ability of homogenizing the temperature across the battery pack [6,11–17]. Therefore, composite phase change materials (CPCM) with thermally highly conductive additives are widely considered to improve the heat transfer [14]. Among various solutions, expanded compressed graphite flakes [18–20] and metallic foams [3,12,13,21–23] have shown promising performances. Particularly, copper foams have shown to increase thermal conductivity up to 5.2 W/mK and the heat transfer rate by 36% [13,21].

It is worth mentioning that the modeling of battery electrochemical response is inevitable in order to develop cost-effective battery packs with efficient thermal management systems. Development of accurate models can substantially decrease the required expensive experimental prototypes and can lead to design more efficient packs as well. Newman's electrochemical pseudo two-dimensional model has been widely employed to simulate the electrochemical response and subsequently evaluating the heat production during charging/discharging cycles [24–26]. Apart from the battery modeling, the complete simulation of a battery pack requires the modeling of surrounding CPCM, which can be simulated with finite element analysis. Nonetheless, efficient FEM simulations ask for the rather complicated modeling and access to an expensive commercial package for the simulations. In order to further enhance the efficiency and decrease related costs, new emerging methods should be explored. With the recent boom of using machine learning algorithms in nearly all industries for tasks like image processing, quality control, predictive maintenance, logistics and autonomously driving vehicles, elaborating a possibility of including those into battery modeling is of particular interest in this work. Machine learning algorithms are statistical computer programs that try to solve a task based on experience from learning. The algorithm learns, if its performance towards the task improves with experience [27,28].

In this work we propose the development of a convolutional neural network to efficiently solve the task of predicting the thermal conductivity of a CPCM for a given image section of the material. This bypasses modeling and computation by finite element analysis and thus save computational resources and modeling costs. Convolutional neural networks (CNNs) are machine learning algorithms which have been proven to be successful in solving image recognition tasks. Recently, they were used to quantitatively predict mechanical properties from images [29,30]. The training data is created by modeling an extensive set of composite sections and conducting finite element analysis. The dataset contains two-dimensional images of metal foam-like structures, which were modeled with different amounts of pores in the plane section. The CNN's performance is compared to a popular CNN for image classification tasks, ResNet18 [31]. A ResNet18 network is trained and evaluated with the same datasets as the developed CNN. With the predicted thermal conductivities, a representative volume element (RVE) is built. The RVE is used to build a battery, which consists of battery cells and surrounding CPCM. The batteries' heat production is simulated by Newman's two-dimensional electrochemical model. The effectiveness of the thermal management system is finally assessed for different CPCM microstructures and the overall accuracy of our proposed approach is examined.

## 2. Methodology

As stated in the introduction, the conducted multiscale modeling in this study includes four steps: finite element modeling of CPCMs and creating the training dataset, developing convolutional neural networks, Newman's electrochemical modeling of a battery cell to acquire the heat sources and finally the finite element modeling of a battery pack thermal management with CPCM. In the following the computational details of each modeling process are presented.

### 2.1 Modeling of CPCMs and creating the training dataset

For creating the training datasets, we generate square two-dimensional sections of CPCM, consisting of paraffin as the PCM and Cu-metal foams of varying densities and pore sizes. In Fig. 1, the developed modeling strategy for creating the dataset is illustrated. The main objective of Cu-metal foam is to enhance the thermal conductivity. In this modeling process, the challenging part is to develop representative microstructure of Cu-metal foam. The underlying geometry of the metal foam in this work is constructed using the Voronoi method via Python scripting. This methodology has been widely employed to model naturally occurring phenomena like polycrystalline microstructures, cell structures of plants or in foams made out of soap bubbles [32]. The metal foam is created by applying morphology to the Voronoi ridges as characterized by Jang *et al.* [33]. First, random points are positioned in the defined section, representative of original seeds for Voronoi tessellation (as shown in Fig. 1). The foam morphology is primarily dependent on its nominal cell size and measured in pores-per-inch (ppi). In principle, the smaller the pore sizes, the shorter and narrower the struts of the metal foam become on average. In our method, smaller pore sizes can be constructed by placing more original seeds. In addition, the strut's cross sectional area was found to change along its length [33]. The struts are then merged with the paraffin matrix to form the CPCM as illustrated in Fig. 1. The presented method produces foams with densities varying from 7.4 to 10.5 % and nominal cell sizes of 9 – 22 ppi. It is worthy to note that more accurate three-dimensional modeling of metal foam CPCM can be achieved, as already conducted by Abishek *et al.* [34]. The focus of this work is on predicting the effective thermal properties of a given set of microstructures with a CNN rather than generating microstructures that precisely correspond to real experimental CPCM samples. After generating the microstructures, the effective thermal conductivity of the composite section is calculated using the finite element analysis (FEA). The finite element modeling of CPCM microstructure in this work is performed with ABAQUS CAE/standard. In the modeling of the CPCM, the thermal conductivity of paraffin and Cu-foams are defined as 0.2 W/mK and 393 W/mK, respectively. The thermal conductivities of the resulting CPCM sections are in the range of 2 – 11 W/(mK), as determined by the FEA results. To create the required dataset for the training of the CNN, an automation algorithm is developed to generate random microstructures, apply boundary conditions and finally evaluate the effective thermal conductivity using the python scripting in conjunction with ABAQUS package. As schematically shown in Fig. 1, one data point of the resulting dataset consists of the effective mean values of the thermal conductivity of the CPCM and the corresponding image of the two-dimensional model. The created dataset in this work includes 20000 images and corresponding thermal conductivity values. The images are in RGB format made of 442x442 pixels.

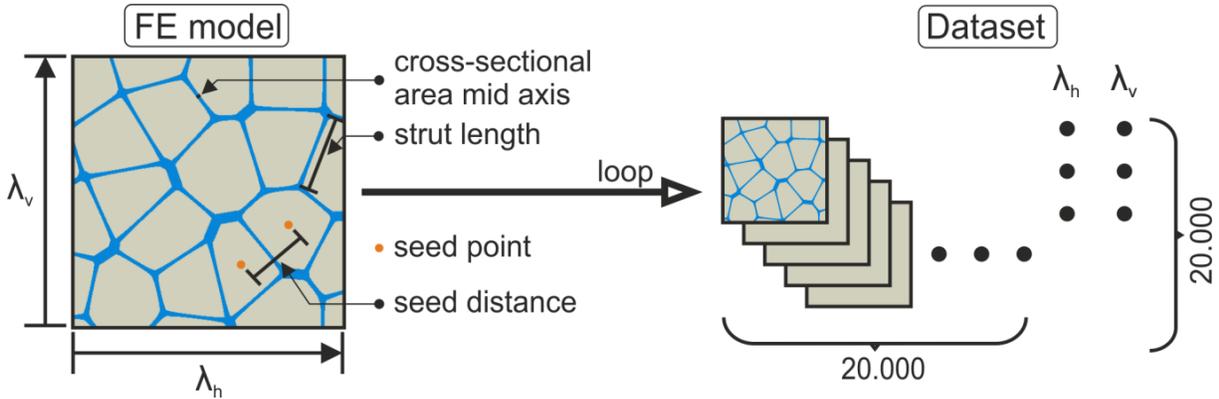

Fig. 1, The developed modeling strategy for creating the dataset of composite phase change materials.

## 2.2 Development of the convolutional neural network

In the conducted CNN modeling, an image acts as the input (x) and the corresponding thermal conductivity is the output or label (y). The neural network's task is to perform regression and predict the labels from the input image. Prior to feeding into the CNN model, the images are pre-processed by resizing to 384x384x3 pixels. Additionally, the pixels of each image are normalized by its mean value and standardized by its standard deviation for each of the three RGB channels. We have chosen CNNs as they provide robust solution for image processing tasks. The architecture of the CNN is created within PyTorch framework [35]. As illustrated in Fig. 2, the constructed neural network contains six layers with learnable weights. Layers 1-4 are convolutional layers, while layers 5 and 6 are fully connected layers, also called dense layers. The output of the last fully connected layer is a tuple, consisting of the predicted horizontal and vertical thermal conductivities. A basic convolutional layer consists of the actual convolutional operation, a non-linear activation function (rectified linear unit "ReLU") and a max-pooling layer. The max-pooling layer effectively performs downsampling of the image width and height by factor 2. Due to this, it is important that the image size is divisable by two for several times. In practice, batch normalization layers have proven effective to speed up learning and lower the generalization (or testing) loss. Therefore, within a convolutional layer, a batch normalization layer is implemented after the convolution operation and before the ReLU non-linearity. To the layers that perform convolution, layers of zeroes are added to preserve the images size during the convolution operation before passing its output to the max-pooling operation. This technique, known as padding, prevents output size reduction and the loss of information in the corners of an image. To prevent overfitting of the network, the first fully connected layer is followed by a dropout layer with a probability of 0.5. In practice, the dropout layer is an effective method for further decreasing generalization loss. The dataset of 20000 images is divided into a training set and a test set with a distribution of 90% to 10%. The network learns by updating its filters. The properties that define the CNN and guide the sampling of the dataset and updating of weights are called hyperparameters. In total the network consists of 434111 learnable weights or parameters. The number of epochs and the batch size are each selected with 4 for both parameters. In this work, the optimizer used is AdamW with default parameters from the PyTorch framework [35]. The learning rate, weight decay and epsilon are 0.001, 0.01 and 0.8 respectively. The constructed microstructure images are fed into the network in mini-batches of four images and associated labels. While sampling, shuffling of the training set is applied to prevent the network from learning image sequences throughout the epochs and get stuck in local minima. In the forward step, the model calculates an output tensor from the input. The mean-squared-error (MSE) is used as loss function for determining the error between predicted outcome of the mini-batch and its labels (y):

$$MSE = \frac{1}{n}\sum_{i=1}^{n}(y_i - \hat{y}_i)^2 \qquad (1)$$

After training is completed, testing is conducted to evaluate the final generalization performance of the model.

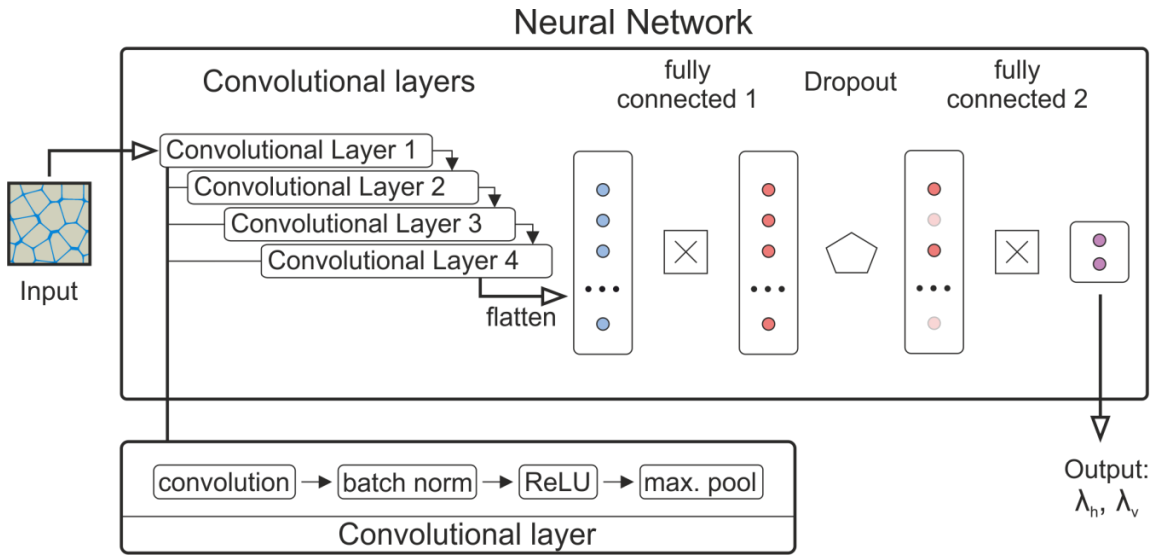

Fig. 2, Architecture of the developed convolutional neural network.

## 2.3 Electrochemical modeling of a battery cell

A typical Li-ion battery consists of two current collectors, cathode and anode electrodes and the electron-insulating separator. The cathode and anode electrodes and separator are porous structures filled by an electrolyte for Li-ion transfer. As proposed by the Newman's pseudo 2D model, the electrochemical response of a Li-ion cell can be evaluated by solving four partial differential equations (PDE) as listed in Table 1 [25,26,36–42]. In this approach the time dependent potentials and Li-ion concentration profiles inside the electrode and electrolyte phases can be obtained under charge conservation. For the solution of these equations, three PDEs can be modeled using a 1D model which is coupled to a 2D model to describe the radius-dependent Li-ions diffusions inside the active particles as shown in Fig. 3.

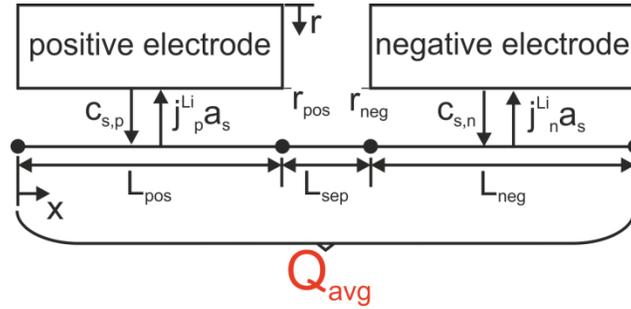

**Fig. 3**, Schematic of battery modeling: 1D electrochemistry model (3 PDEs), 2D electrode model (1 PDE).

In our study, the anode and cathode active materials are assumed to be the meso-carbon micro-bead (MCMB) graphite and $LiCoO_2$, respectively [38]. The four governing differential equations and their corresponding boundary conditions are summarized in Table 1. Since we deal with multiphase structures, we used the Bruggeman's relation [43,44] to evaluate the effective medium properties as also summarized in Table 1. The ion transfer at the interface between solid particles and electrolyte is simulated using the Butler-Volmer equation as follows:

$$j^{Li} = a_s i_0 \left\{ \exp\left[\frac{0.5F}{RT}\eta\right] - \exp\left[-\frac{0.5F}{RT}\eta\right] \right\} \quad (2)$$

Here $a_s$ is the particles' surface area per unit volume and $i_0$ is the exchange current density defined via:

$$i_0 = K_i (C_e)^{0.5} (C_{s,max} - C_{s,surf})^{0.5} (C_{s,surf})^{0.5} \quad (3)$$

$K_i$, $C_{s,max}$ and $C_{s,surf}$ are, respectively, reaction rate coefficient and maximum and surface content of Li in the active particles, and $C_e$ is the Li-ion concentration in the electrolyte. The driving force for the reaction is the over-potential $\eta$, which is expressed by:

$$\eta = \phi_s - \phi_e - U \quad (4)$$

Here $U$ is the experimentally measured equilibrium potential of active materials as a function of intercalated Li. The equilibrium potentials for the MCMB graphite and $LiCoO_2$ as a function of Li-ion concentration are taken from the work by Kumaresan *et al.* [38]. The properties of the electrolyte LiPF6 in EC:EMC (3:7), in this case the diffusion coefficient, its ionic conductivity and the lithium ion transport number, were characterized by Nyman *et al.* [45] and also vary strongly with lithium concentration. The corresponding electrochemical parameters used in this work are listed in Table 2. Based on the presented data the 1C current density for this cell is calculated to be 20 A/m². COMSOL commercial package is employed for the implementation of Newman's pseudo 2D model via using a 1D model with 3 line segments for respectively, describing the negative electrode, separator, and positive electrode. The potential in the active materials and electrolyte, the Li-ion content in the electrolyte are simulated using the 1D model. In this model the reaction current density is also obtained for the cathode and anode domains. The Li-ion inside the active material is however simulated using the rectangles, in which the width is representative of particle's average radius. The acquired reaction current density from the 1D solution is then projected to define the boundary condition along the length of aforementioned rectangle [40,41] (see Fig. 3). The implemented Newman's model in this study has been validated in previous works [40,46]. Worthy to note that for the sake of simplicity in this study the effects of temperature change on the electrochemical response is neglected.

Next we can evaluate the heat generation in the Li-ion cell, which consists of three main sources: heat from the reaction current and over-potentials ($Q_r$), ionic ohmic heat from the carried current inside the active materials and electrolyte phases ($Q_j$), and heat due to the entropy changes in active materials ($Q_{rev}$), which are evaluated, respectively, as follows:

$$Q_r = Aj^{Li}(\phi_s - \phi_e - U) \quad (5)$$

$$Q_j = \sigma^{eff}\left(\frac{\partial \phi_s}{\partial x}\right)^2 + \kappa^{eff}\left(\frac{\partial \phi_e}{\partial x}\right)^2 + \kappa_D^{eff}\frac{\partial(\ln C_e)}{\partial x}\left(\frac{\partial \phi_e}{\partial x}\right) \quad (6)$$

$$Q_{rev} = j^{Li}T\frac{\partial U}{\partial T} \quad (7)$$

We note that unlike the other two source of heat, the heat from entropy changes is reversible and can take positive and negative values depending on the $\partial U/\partial T$ relations. Here, $\partial U/\partial T$ curves as a function of Li content for MCMB graphite and LiCoO$_2$ are taken from the work by Kumaresan *et al.* [38].

**Table 1**, Governing equations and boundary conditions according to the Newman's pseudo-2D model. The effective properties were obtained mainly by using the Bruggeman's relation.

| **Electrochemical equations** | | **Boundary conditions** |
|---|---|---|
| Mass balance, electrolyte phase | $\frac{\partial(\varepsilon_e c_e)}{\partial t} = \frac{\partial}{\partial x}\left(D_e^{eff}\frac{\partial}{\partial x}c_e\right) + \frac{1-t_+^0}{F}j^{Li}$ | $\frac{\partial c_e}{\partial x}\|_{x=0} = \frac{\partial c_e}{\partial x}\|_{x=L_n+L_{sep}+L_p} = 0$ |
| Mass balance, solid phase | $\frac{\partial c_s}{\partial t} = \frac{D_s}{r^2}\frac{\partial}{\partial r}\left(r^2\frac{\partial c_s}{\partial r}\right)$ | $\frac{\partial c_s}{\partial r}\|_{r=0} = 0, -D_s\frac{\partial c_s}{\partial r}\|_{r=R_s} = \frac{j^{Li}}{a_s F}$ |
| Electrical potential, electrolyte phase | $\frac{\partial}{\partial x}\left(\kappa^{eff}\frac{\partial}{\partial x}\varphi_e\right) + \frac{\partial}{\partial x}\left(\kappa_D^{eff}\frac{\partial}{\partial x}\ln c_e\right) + j^{Li} = 0$ | $\frac{\partial \varphi_e}{\partial x}\|_{x=0} = \frac{\partial \varphi_e}{\partial x}\|_{x=L_n+L_{sep}+L_p} = 0$ |
| Electrical potential, solid phase | $\frac{\partial}{\partial x}\left(\sigma^{eff}\frac{\partial}{\partial x}\varphi_s\right) = j^{Li}$ | $-\sigma_n^{eff}\frac{\partial \varphi_s}{\partial x}\|_{x=0} = \sigma_p^{eff}\frac{\partial \varphi_e}{\partial x}\|_{x=L_n+L_{sep}+L_p} = \frac{I}{A}$ |
| | | $\frac{\partial \varphi_s}{\partial x}\|_{x=L_n} = \frac{\partial \varphi_s}{\partial x}\|_{x=L_n+L_s} = 0$ |
| **Effective properties** | | |
| Electrolyte ionic diffusivity | | $D_e^{eff} = D_e \varepsilon_e^{brug}$ |
| Electrolyte ionic conductivity | | $\kappa^{eff} = \kappa \varepsilon_e^{brug}$ |
| Electrolyte ionic diffusion conductivity | | $\kappa_D^{eff} = \frac{2RT\kappa^{eff}}{F}(1-t_+^0)$ |
| Solid phase electronic conductivity | | $\sigma^{eff} = \varepsilon_s \sigma$ |

Table 2, Parameters of the Li-ion battery cell modeling.

| Parameter | Description | Negative electrode | Separator | Positive electrode |
|---|---|---|---|---|
| $t^0$ | Transference number of electrolyte [45] | 0.3 | 0.3 | 0.3 |
| $L$ | Length of electrode (μm) [47] | 75 | 28 | 75 |
| $r$ | Solid particle radius (μm) [48,49] | 5 | — | 3 |
| $\varepsilon_s$ | Solid phase volume fraction [50] | 0.4 | — | 0.419 |
| $\varepsilon_e$ | Electrolyte phase volume fraction [50] | 0.45 | 0.55 | 0.413 |
| $\varepsilon_b$ | Binder volume fraction [50] | 0.15 | — | 0.168 |
| $D_s$ | Solid phase diffusion coefficient (m$^2$/s) [39] | 3.89 × 10$^{-14}$ | — | 1 × 10$^{-13}$ |
| $P$ | Active material density (kg/m$^3$) [39] | 2292 | — | 5032 |
| $C_{s,max}$ | Maximum concentration in solid phase (mol/m$^3$) [38,51] | 31 370 | — | 49 943 |
| $SOC_0$ | Initial state of charge in Charge/Discharge [38,48] | 0.84/0.01 | — | 0.44/0.97 |
| $C_{e,0}$ | Initial electrolyte concentration (mol/m$^3$) [38,39,51] | 1000 | 1000 | 1000 |
| $brug$ | Bruggeman coefficient for tortuosity [25] | 1.5 | 1.5 | 1.5 |
| $K_i$ | Reaction rate coefficient (A m$^{2.5}$/mol$^{1.5}$) [38] | 1.764×10$^{-11}$ | — | 6.67×10$^{-11}$ |
| $\sigma$ | Solid phase conductivity (S/m) [25,38,52] | 100 | — | 10 |

## 2.4 Finite element modeling of a battery thermal management system

After evaluating the generated heat sources $Q_{avg}$ during the battery charge/discharge operations, we are finally able to examine a battery thermal management system. The considered battery module consists of 6 Li-ion batteries inside CPCM with a rectangular layout. The battery pack dimensions are shown in detail in Fig. 4. Thermal properties of Li-ion batteries are taken from the work by Sabbah *et al*. [53]. Accordingly, the thermal conductivity of the battery cell is 3 W/mK and 30 W/mK in radial and axial direction, the heat capacity is fixed at 900 J/kgK and the density is assumed to be 2663 kg/m$^3$. Thermal conductivity and heat capacity of CPCMs evaluated previously are used to define corresponding properties in the FEM modeling [46]. The melting temperature range of the paraffin is within 41 - 49 °C, with peak phase change temperature at 46.7 °C. The latent heat of considered paraffin is 228 J/g [54]. We note that the density is calculated by the rule of mixtures. Since the latent heat flux of paraffin varies considerably with the temperature, the latent heat of CPCM in our FEM modeling is defined as a temperature dependent heat capacity function. The temperature dependent heat capacity of paraffin and copper in this work are taken from the work by Ukrainczyk *et al.* [54] and Shomate equation [55], respectively. The resulting heat capacity of the CPCM is defined using the following relation [46]:

$$C_{p,CPCM}(T) = \frac{C_{p,Cu}(T)\varepsilon_{Cu}\rho_{Cu} + C_{p,Pa}(T)\varepsilon_{Pa}\rho_{Pa}}{\varepsilon_{Cu}\rho_{Cu} + \varepsilon_{Pa}\rho_{Pa}} \qquad (8)$$

Here the *Cp*, *ε* and *ρ* are, respectively, the heat capacity, volume fraction and density of paraffin or Cu-foam. The volumetric heat sources from the electrochemical solution are finally applied to the volumes of the battery cells. The heat transfer to the ambient air is simulated by defining a convective heat transfer coefficient of 5 W/(m$^2$K) over all outer surfaces of the constructed model shown in Fig. 4.

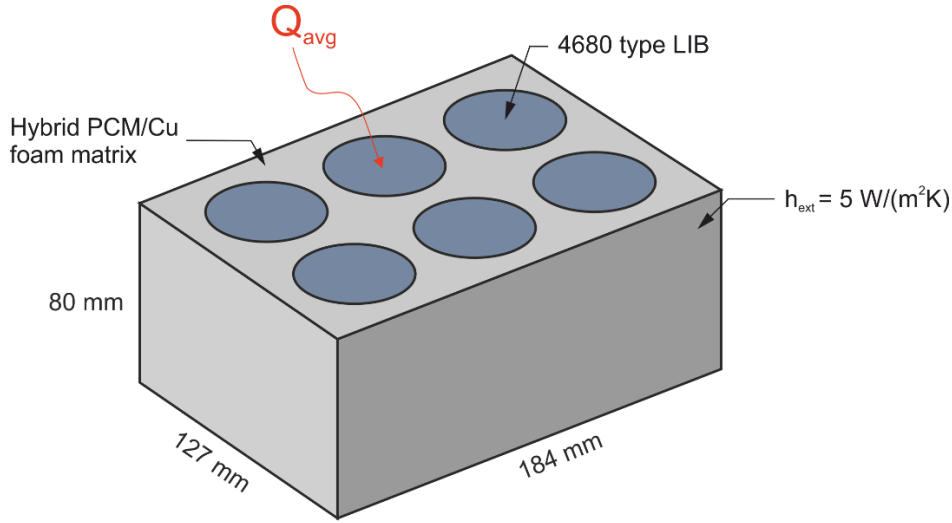

Fig. 4, Developed heat transfer model to examine the thermal management of a battery pack made of 6 Li-ion cells embedded in the composite phase change material. The dimeter and heights of each battery are 46 and 80 mm, respectively. In this modeling strategy the average volumetric heat generation ($Q_{avg}$) from the electrochemical solution is used to define the heat source. The outer surfaces of this are in contact with air simulated via convective heat transfer.

## 3. Results and discussions

In this section, we first evaluate the performance of developed CNN. During the training process, an optimization task is performed in order to minimize the loss function of training by adapting the weights of the CNN. The results are compared with results achieved with a CNN model with ResNet18 architecture. Worthy to note that residual networks (ResNets) usually omit dropout layers. In result, ResNets allow the usage of more layers than plain networks and possibly achieve better performance, while plain networks start to perform worse when using too many layers[31]. ResNet18 is a predefined and versatile network that can be employed for a wide variety of problems. This approach performs identity mapping from a previous layer to the next. As the chosen number of epochs is four, each image of the dataset is fed four times into the CNN. The loss convergence histories from both developed CNN and ResNet18 are presented in Fig. 5. The training and testing loss curves show expected behavior, where the testing loss is always slightly greater than the training loss. Both networks show comparable performances and can predict the thermal conductivity of the composite sections with high accuracy. The minimum testing losses (MSE) with respect to the mean thermal conductivity of the dataset are 3.5 % and 3% over the whole test set for the developed CNN and ResNet18, respectively. These accuracies are calculated by evaluating the images from the test set that are not utilized during training. The small difference between training and test loss indicates good generalization for both network architectures. The learning speed of both models is also found to be comparable, with almost two epochs required for reaching a plateau in testing loss.

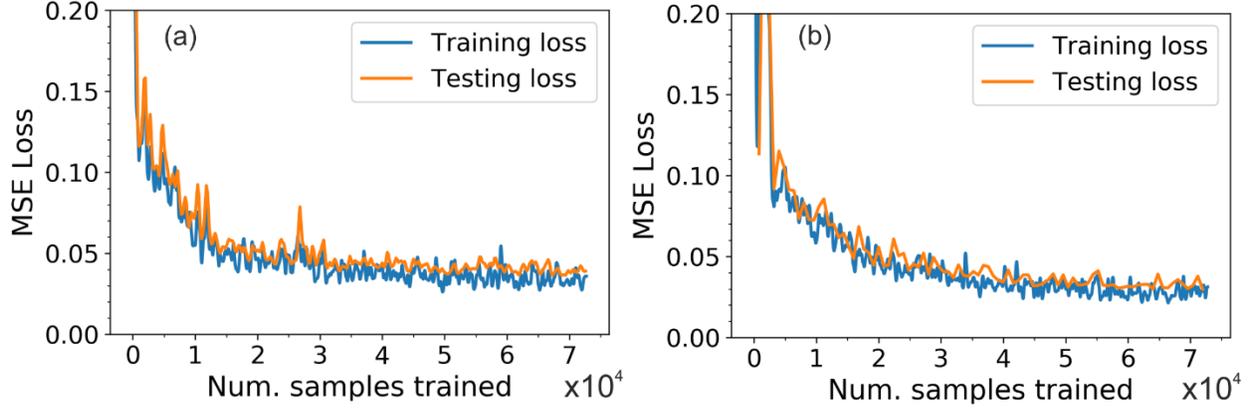

**Fig. 5**, Convergence history of loss functions with respect to the average thermal conductivity of the dataset for training and testing of (a) own developed CNN and (b) ResNet18.

In order to better examine the performance of developed neural networks, the dataset is classified into different ranges of thermal conductivity with corresponding number of samples during training. For each class, the mean absolute percentage error $MAPE = \frac{1}{n}\sum_{i=1}^{n}\left|\frac{y_i-\hat{y}_i}{y_i}\right|$ and the maximum relative testing error are determined and the acquired results are displayed in Fig. 6. Due to the small size of the test set, large fluctuation can be observed on the maximum relative errors. First, the predictions of horizontal thermal conductivity are analyzed. The maximum errors are in the range of 19 – 55 % and 11 – 61 % for the developed CNN and ResNet18, respectively. As expected, for the samples with either low or high thermal conductivities, higher average losses occur, because they are rarely sampled. While the maximum errors provide unique understanding about worst performance of the developed models, the mean absolute percentage error is a more representative measure of the overall performance and decreases as the number of samples used during training increases. Another representative performance measure is the portion of predictions that have a mean absolute percentage error greater than 10%. For this evaluation, the most frequently sampled thermal conductivity classes during training are taken into account. For the developed CNN, the mean absolute percentage error is 5 % on average while 14 % of predictions have a greater error than 10%. In the case of the ResNet18 model, the mean absolute percentage error is also 5 % with the corresponding portion of predictions with error greater than 10 % is 11.4 %.

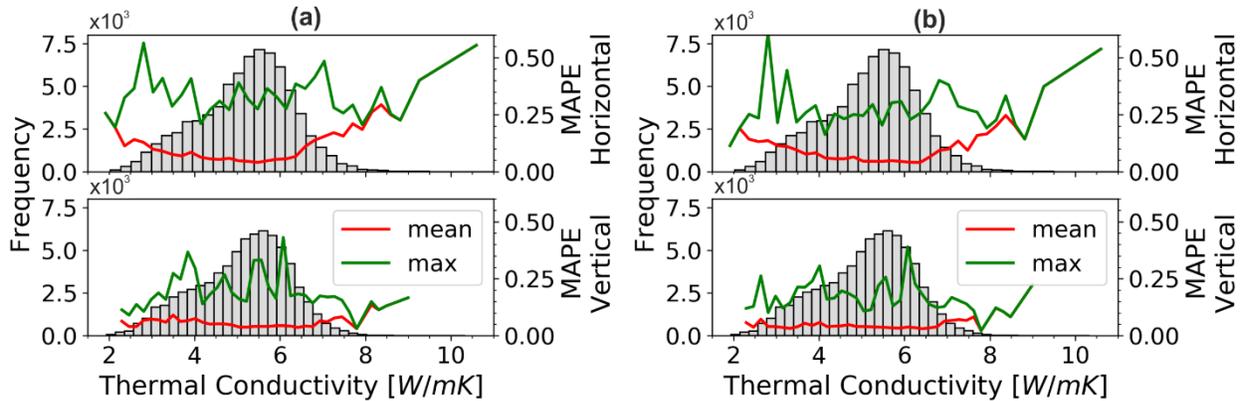

**Fig. 6**, Frequency of sampled thermal conductivity ranges and relative loss for a range of thermal conductivity for (a) developed CNN and (b) ResNet18.

The predictions for vertical thermal conductivity are in general slightly more accurate in terms of mean and maximum errors for both network architectures. According to the results shown in Fig. 6, the maximum errors are in a range of 3 – 43 % and 2 – 39 % for the developed CNN and ResNet18, respectively. The mean absolute percentage error is 4.2 % for the developed CNN and 3.9 % for ResNet18. The developed CNN predicts only 9 % of samples with a higher relative error than 10 %. The evaluation with ResNet18 shows that only 3.2 % of the predictions have a higher mean absolute percentage error than 10 %. From these results it is clear that ResNet18 slightly outperforms developed CNN with respect to the accuracy, which can be associated with higher number of weights in ResNet18. From the results shown in Fig. 6, it is clear that the difference between the horizontal and vertical conductivities arise for the images with either low or high thermal conductivities. These samples are less explored in the training procedure and therefore they show highest errors. The difference between horizontal and vertical thermal conductivities is expected to disappear with a larger dataset or by shuffling the dataset from a different seed. In addition, from the modelling point of view, by considering very large FEM models the difference between horizontal and vertical thermal conductivities is expected to vanish. Nonetheless, by considering large RVEs not only the computational cost increases considerably but also the feature recognition by the CNN approach become more complex. The good agreement between the predictions from the developed CNN model and the finite-element-data shows the high potential of CNN models to be used in determining effective material properties. Especially the low mean absolute percentage errors allow applications where several cross sections of a composite material are used to determine the thermal properties.

We next shift our attention to the electrochemical results. Here we study the cell voltage and average volumetric heat sources generated during the battery operation. The voltages are time-dependent functions and are presented in Fig. 7 for charging and discharging and various current densities with respect to the 1 C current. During the simulation, the cell voltage can be obtained by subtracting the potential of the anode from the potential of the cathode: $V = \Phi_s|_{x = Ln+Lsep+Lp} - \Phi_s|_{x = 0}$. For charging or discharging the battery, the initial states of charge are listed in Table 2. At the beginning of charging or discharging, a jump in the cell potential can be observed, which represents the increasing of overpotential to reach the chosen current. As expected, the jump increases with increasing current density. After removing the load from the battery, a relaxation effect occurs and the potentials converge towards an equilibrium value. As reaction rate coefficients differ from anode to cathode, the initial jump in the voltage curve of charging is significantly more pronounced than for the profile of discharging.

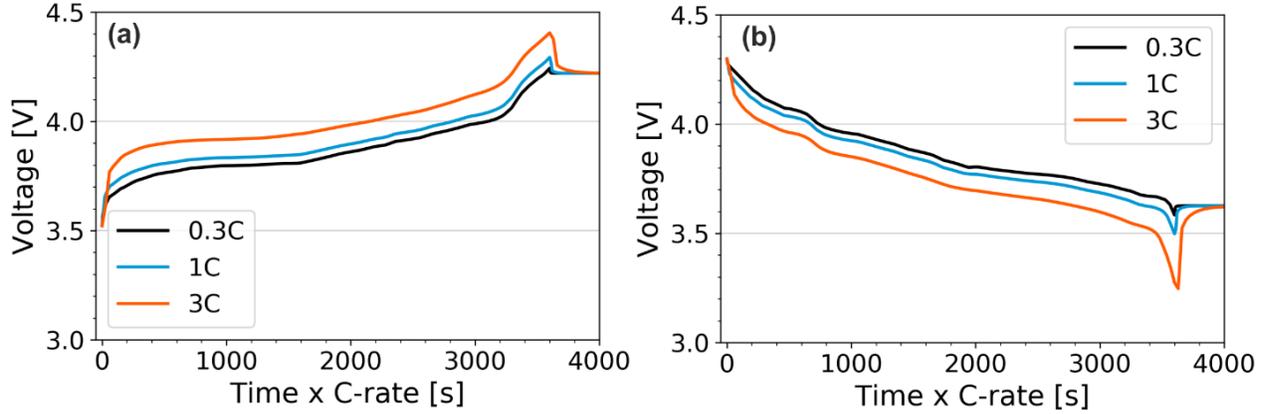

Fig 7, Simulated cell voltage curves as a function of time for (a) charging and (b) discharging cases with different current densities with respect to 1 C current (20 A/m$^2$).

The average volumetric heat generated during charging and discharging is calculated for 0.3 C, 1 C and 3 C current densities, as illustrated in Fig. 8. The total heat generated during the battery operation can be determined using the energy balance equation. An obvious characteristic is the higher magnitude of heat generation for the discharging cycle compared to the charging cycle. To better understand the underlying mechanism for such a significant difference, we should take the reversible heat generation into account. The reversible heat from Eq. 7 transferred to anode and cathode is $j^{Li}T(\frac{\partial U_p}{\partial T} - \frac{\partial U_n}{\partial T})$ and can be either positive or negative depending on the states of charge of anode and cathode and current direction. ∂U/∂T curves as a function of the state of charge for anode and cathode are taken from experimental measurements [38] and show irregular patterns. In our considered battery model, consisting of graphite anode and a LiCoO$_2$ cathode, the ∂U/∂T curve of the cathode shows greater magnitudes over the state of charge and thus this electrode contributes to the majority of reversible heat generation during operation. For the case of discharging, the lithium concentration in the cathode is initially low and the ∂U/∂T profile is negative, which results in the increased heat generation, as also observable from results shown Fig. 8 (b) and (d). By further increasing of the state of the charge in cathode during the discharge, the ∂U/∂T profile increases to positive range, which results in cooling. The remaining discharging process only generates heat, which increases significantly when the cathode is close to the fully charged state. As observable from results shown in Fig. 8, opposite trends occur for the charging cycle. It can also be seen that unlike the reversible heat generation, the irreversible heat initially jumps at the beginning after applying the load and keeps an almost constant value for the remaining time. This observation is expectable since the irreversible heat is of an ohmic nature and is proportional to the current density with power of two. Acquired results in accordance with previous studies [38,46] highlight the importance of the reversible heat generated in the LiCoO$_2$ cathode. However, as the current density increases, the contribution of the reversible heat to the total heat decreases and heat generation is more dominated by the irreversible heat, because the reversible heat scales linearly with the applied current, whereas the irreversible heat scales quadratically with the current density.

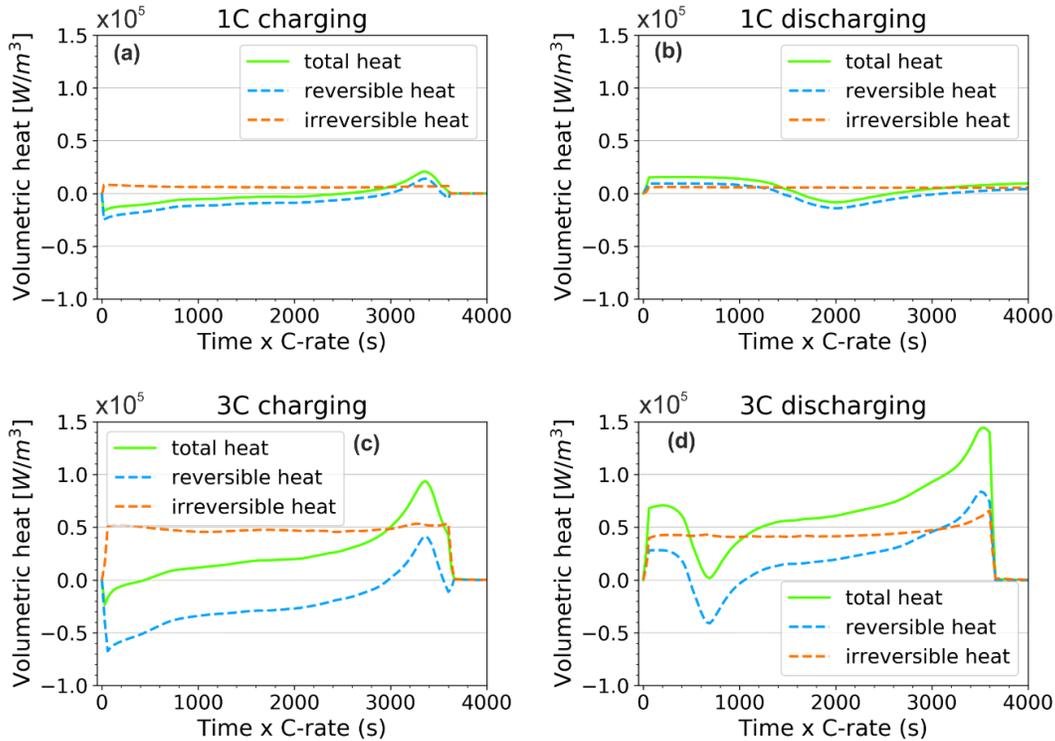

**Fig. 8**, Simulated average volumetric heat sources as a function of time for cases (a) 1C charging, (b) 1C discharging, (c) 3C charging and (d) 3C discharging.

We now investigate the impact of different charging and discharging currents on the temperature rise of the considered battery pack. We first assume the thermal conductivity of the CPCM as the mean value of all constructed FEM models, equal to 5.2 W/mK. From the earlier obtained volumetric heat generation rates, the battery pack's temperature is expected to be higher for higher currents and for discharging in comparison with the charging cycle. The results for the maximum temperature rise in the battery pack illustrated in Fig. 9 confirm the aforementioned expectations. In this regard, the temperature increases more with the 4 C current compared to the 1 C current, due to the increases ohmic heat generation. Moreover, with a fixed current, higher temperatures develop in the system during discharging than during charging. It is noticeable that the temperature rises during the 1 C discharging and 4 C charging are close. For the 4 C case, the temperature increase during the charging cycle is only half of the temperature increase that occurs during discharging. Interestingly for the case of 1 C charging current, the battery does not heat up at all, but is instead cooled especially in the initial stages of charging. Towards the end of the cycle there is a heating up, but the battery temperature does not exceed the ambient temperature. For the 4 C charging case, the temperature barely rises during the initial 200 seconds of loading, whereas a substantial rise is noticeable for 4 C discharging current. These observations are in agreement with results in previous studies [24,38,46]. Since for thermal management the discharge process is more critical than the charging process, the discharging process is further considered.

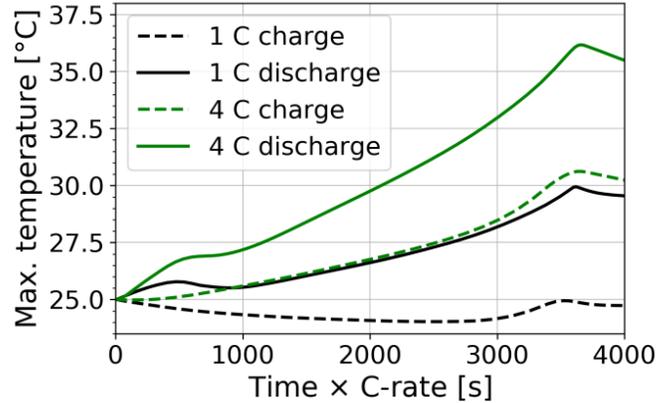

Fig. 9, Simulated maximum temperature rise in the battery pack as a functions simulation time for 1 C and 4 C charging and discharging currents.

Finally, we examine the accuracy of developed CNN in predicting the temperature rise of the considered Li-ion pack as compared with the full-FEM solution. In this regard, two samples of the CPCM will be investigated to simulate the temperature rise within the batteries. We assume that the microstructure of a CPCM sample can be well-represented with 5 cross section images. Here, the considered images were not involved in CNN training. Fig. 10 shows the two CPCM samples and their corresponding cross section images. In the aforementioned illustration we also plotted the surfaces temperature profiles of the battery pack for different times during discharge with 4 C current, computed with the thermal properties of CPCM obtained by FEA and CNN. Note, that earlier, results were plotted over the time x C-rate, while in this case only the real time is represented. The images show the process of heat generation inside the battery cells. By the 920 s timestamp, the maximum temperatures are achieved during the simulation, and therefore the highest temperature gradients from battery to CPCM. One observation is that the temperature inside the CPCM is relatively uniform. The heat conduction from the surface of the battery cells to the ambient air therefore works sufficiently well, which confirms the effectiveness of a highly conductive CPCM. However, noticeable heat accumulation at the center axis of the battery cells is observable. This finding is due to the low thermal conductivity of the battery cells in the radial direction, which is lower than the thermal conductivity of the CPCM. Such that a marginal temperature gradient stablishes inside batteries. At the time stamp 1150 seconds, there is no more discharging load on the battery and in this case the temperature is distributed from the battery to the CPCM and from the CPCM surfaces to the environment. Note that although the thermal conductivity of sample 1 is almost twice of that of the sample 2, the time evolution of maximum temperature rise is yet very close. This can be partially explained by the fact that the thermal conductivity of sample 2 still exceeds the thermal conductivity of the battery, which is one of the limiting factors in conducting the heat from the inside of cells. Please note that by increasing the content of metal foams, although the thermal conductivity increases the heat capacity decreases. Heat capacity and latent heat of the CPCM, the thermal conductivity of the battery itself and the convective heat transfer to the environment via the CPCM surfaces are contributing factors to the maximum temperature rise.

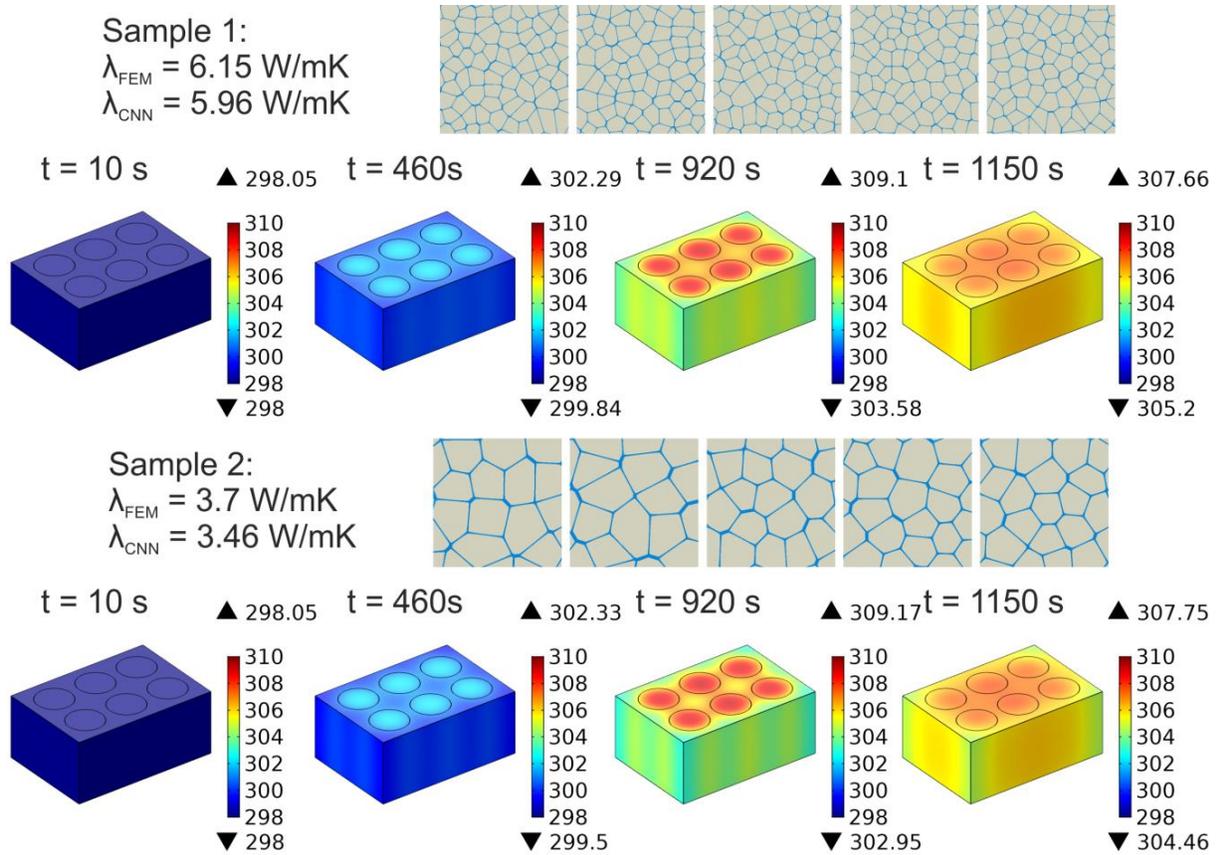

Fig. 10, Constructed two CPCM samples and their corresponding cross section images and estimated thermal conductivities with CNN ($\lambda_{CNN}$) and FEM ($\lambda_{FEM}$). Time evolution of surface temperatures of Li-ion battery pack with two different CPCM samples during the 4 C discharging current.

In order to better examine the accuracy of developed CNN in predicting the temperature rise of the Li-ion pack surrounded by two constructed CPCMs, we plot the difference in the maximum temperature rise using the FEM and CNN method for introducing the thermal properties of CPCM. We conducted the simulations for different charging and discharging currents and the obtained results are depicted in Fig. 11. The maximum absolute temperature differences between CNN and FEM models is found to be maximal for 4 C discharging loading. As observable, the maximum differences for the absolute temperature are below 0.005 % and 0.015 % for the whole discharging cycle for sample 1 and sample 2, respectively. As the maximum cell temperature is the key parameter for evaluating the performance of a thermal management system, our results highlight the remarkable accuracy of CNN and its ability to substitute complex FEM modeling.

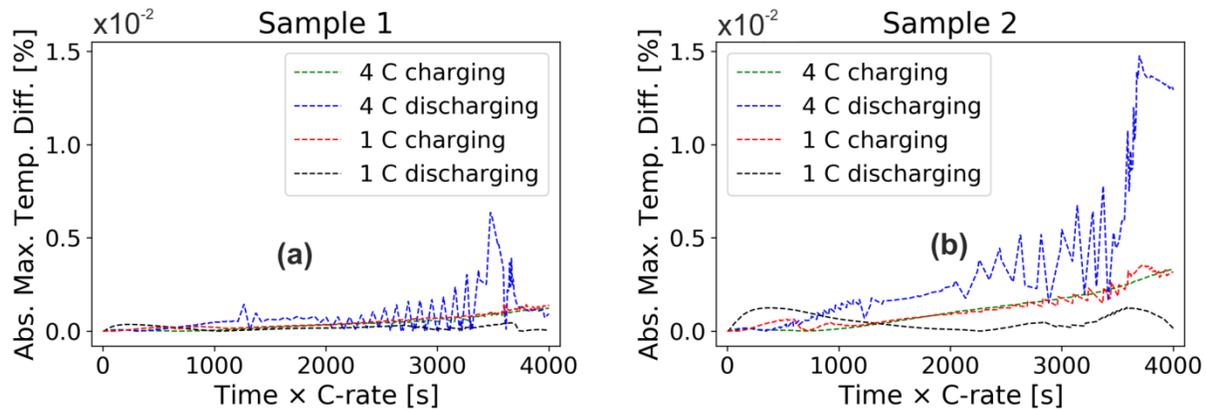
Fig. 11, Absolute maximum temperature difference in % with respect to the absolute temperature as a function of time x C-rate for (a) sample 1 and (b) sample 2.

## 4. Conclusion

In this work, we show that convolutional neural networks (CNNs) trained over finite element method (FEM) simulations can accurately reproduce the thermal conductivity of complex composite phase change materials (CPCMs). The CNN's performance is also compared with that of a ResNet18 model, which is a popular CNN architecture for image recognition tasks. Both neural network models show very similar performance characteristics. We show that a CNN network with a relatively low number of weights trained over small dataset achieves mean absolute percentage errors of around 5%. As a practical example, we examine the accuracy of the developed CNN in predicting the temperature development in a Li-ion pack with passive thermal management using the CPCM, during different charging/discharging currents. For this purpose, an electrochemical modeling of a battery chemistry is conducted and the average volumetric reversible and irreversible heat generation rates are extracted. The results for the case of $LiCoO_2$ cathode show the importance of reversible heat generation during the battery operation. It is also illustrated that the heat generation is larger during discharging than the charging cycle. The 3D heat transfer model results for the effectiveness of the CPCM thermal management system, confirm that the model developed on the basis of CNN can robustly with ultrahigh precision estimate the maximum temperature rise in the battery pack. Our study not only proposes a modeling strategy to facilitate the examination of thermal properties of complex composite structures, but also highlights the accuracy and application prospect of CNNs to substitute complex, time consuming and expensive FEM modeling.

## Acknowledgment

B.M. and X.Z. appreciate the funding by the Deutsche Forschungsgemeinschaft (DFG, German Research Foundation) under Germany's Excellence Strategy within the Cluster of Excellence PhoenixD (EXC 2122, Project ID 390833453). Authors also acknowledge the support of the cluster system team at the Leibniz Universität of Hannover.

## Nomenclature

| | |
|---|---|
| $a_S$ | active surface area per electrode unit volume ($m^2/m^3$) |
| $c$ | concentration of Li in a phase ($mol/m^3$) |
| $C_p$ | heat capacity of a phase (J/gK) |
| $V$ | cell voltage (V) |
| $D$ | diffusion coefficient of lithium species ($m^2/s^1$) |
| $F$ | Faraday's constant, 96,487 C/$mol^1$ |
| $h$ | convective heat transfer coefficient (W/$m^2$ $K^1$) |
| $i_0$ | exchange current density of an electrode reaction (A/$m^2$) |
| $I$ | applied current (A) |
| $S$ | contact surface ($m^2$) |
| SOC | state of the charge |
| $j^{Li}$ | reaction current density (A/$m^3$) |
| $K$ | rate constant for electrochemical reaction of electrode (A $m^{2.5}$/$mol^{1.5}$) |
| $L$ | domain's length (µm) |
| $r$ | radial coordinate along active material particle (cm) |
| $R$ | universal gas constant, 8.3143 J/mol K |
| $R_S$ | average radius of solid particles (m) |
| $t_+^0$ | transference number of Li-ion inside electrolyte |
| $T$ | absolute temperature (K) |
| $U$ | open-circuit voltage of an electrode reaction (V) |

### Greek symbols

| | |
|---|---|
| $\varepsilon$ | volume fraction of a phase |
| $\eta$ | surface over-potential of an electrode reaction (V) |
| $\kappa$ | ionic conductivity of electrolyte (S/m) |
| $\kappa_D$ | diffusional conductivity of a species (A/m) |
| $\sigma$ | conductivity of solid particles (S/m) |
| $\varphi$ | electrical potential in a phase (V) |
| $\rho$ | density (kg/$m^3$) |

### Subscripts

| | |
|---|---|
| i | anode, separator, or cathode region |
| e | electrolyte phase |
| s | solid phase |
| b | binder for solid particles in the electrode |
| max | maximum concentration |
| ref | with respect to a reference state |
| s,surf | Li concentration on the surface of particles |
| sep | separator region |
| avg | average |
| n | negative electrode region |
| p | positive electrode region |
| pa | paraffin |
| Cu | Cu-foam |

### Superscripts

| | |
|---|---|
| eff | effective property of a medium |
| Li | lithium species |
| brug | Bruggeman effective medium factor |